\documentclass[sigconf,nonacm]{acmart}

\usepackage{url}            % simple URL typesetting
\usepackage{booktabs}       % professional-quality tables
\usepackage{amsfonts}       % blackboard math symbols
\usepackage{microtype}      % microtypography
\usepackage{xcolor}         % colors
\usepackage{graphicx}
\usepackage{enumitem}
\usepackage{multirow}
\usepackage{amsmath}
\usepackage{mathtools}
\usepackage{tcolorbox}
\usepackage{xspace}
\usepackage{caption}
\usepackage{subcaption}
\usepackage{multirow}
\usepackage{makecell}
\usepackage{xspace}

\newcommand{\our}{\textsc{TTP-R1}\xspace}

\begin{document}

\title{
Retrieval-Constrained Policy Optimization for Attack Technique Extraction from Cyber Threat Intelligence
}

\author{Jiayun Zhang}
\email{jiayunz@amazon.com}
\affiliation{%
  \institution{Amazon Web Services}
  \city{New York}
  \state{New York}
  \country{USA}
}

\author{Junshen Xu}
\email{jsxu@amazon.com}
\affiliation{%
  \institution{Amazon Web Services}
  \city{Boston}
  \state{Massachusetts}
  \country{USA}
}

\author{Zejun Xie}
\authornote{Work done during internship at Amazon Web Services.}
\email{zx180@cs.rutgers.edu}
\affiliation{%
  \institution{Rutgers University}
  \city{Piscataway}
  \state{New Jersey}
  \country{USA}
}

\author{Yi Fan}
\email{fnyi@amazon.com}
\affiliation{%
  \institution{Amazon Web Services}
  \city{New York}
  \state{New York}
  \country{USA}
}

\begin{abstract}
Mapping cyber threat intelligence (CTI) text to MITRE ATT\&CK techniques is essential for structured threat analysis, yet manual annotation is costly and does not scale.
The ATT\&CK taxonomy comprises several hundred attack techniques, and a single CTI passage may describe multiple techniques, making accurate and complete extraction challenging. 
Existing automated approaches fall short in different ways: multi-label classifiers struggle with severe class imbalance and the large label space, while LLM-based methods---retrieval pipelines and fine-tuned generators---optimize token-level objectives that treat technique annotation as sequence generation rather than set prediction, lacking direct supervision on whether the predicted technique set is correct and complete.
We propose \our, a two-stage framework that combines retrieval-augmented supervised fine-tuning (SFT) with reinforcement learning using verifiable rewards (RLVR).
A hybrid retriever first narrows the large label space to a candidate set, and a fine-tuned LLM learns to select the correct techniques.
We then apply Group Relative Policy Optimization with a decomposed reward that directly supervises the precision, recall, and output format of the predicted technique set. 
Across four CTI benchmarks, \our achieves the best average F1, improving sub-technique-level F1 by 7.4 percentage points over Claude Sonnet 4.5 with retrieval augmentation, while running 28$\times$ faster when served as an 8B-parameter model on a single GPU.
\end{abstract}




\maketitle

\section{Introduction}
\begin{figure}[t]
  \centering
  \includegraphics[width=\linewidth]{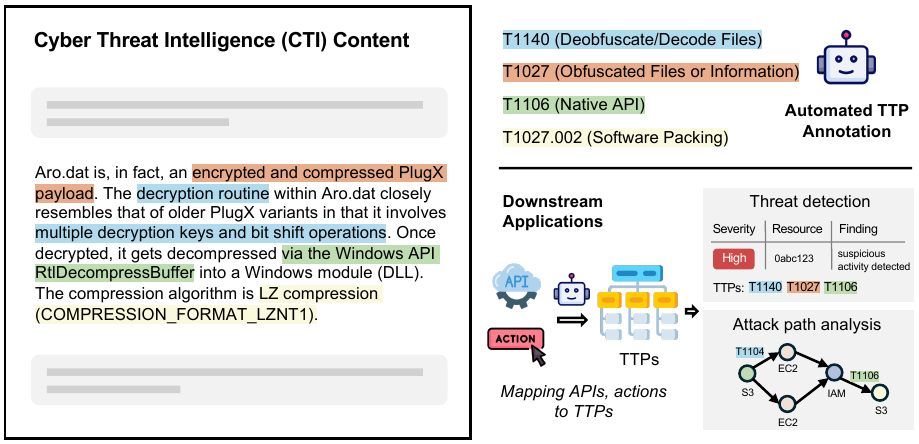}
  \caption{Example of attack technique extraction from Cyber Threat Intelligence (CTI). A CTI passage describing adversary behavior is mapped to MITRE ATT\&CK techniques, including parent techniques (e.g., T1140) and sub-techniques (e.g., T1027.002). The extracted techniques feed downstream applications such as threat detection and attack path analysis.}
  \label{fig:illustration}
\end{figure}

Cyber threat intelligence (CTI) reports describe adversarial behaviors in natural language---the tools, methods, and objectives employed by attackers.
While rich in detail, these unstructured narratives are difficult to operationalize at scale.
To convert them into structured, machine-readable representations, security analysts map CTI text to the MITRE ATT\&CK framework~\cite{strom2018mitre}, a standardized taxonomy that organizes adversary behaviors into \emph{tactics} (high-level goals such as Lateral Movement), \emph{techniques} (specific methods to achieve those goals), and \emph{procedures} (real-world implementations of techniques).
The taxonomy defines over five hundred techniques and sub-techniques, organized hierarchically.
The resulting structured annotations enable downstream applications such as threat hunting~\cite{xu2026multi}, detection rule generation~\cite{shabtai2024llmcloudhunter}, and intelligence sharing~\cite{lekssays2025stix}, and are widely used in production security systems.
Figure~\ref{fig:illustration} gives an example of technique mapping from CTI and downstream applications.
As the volume of CTI publications grows, manually annotating techniques becomes a bottleneck because it requires both breadth of knowledge and fine-grained reasoning, and cannot scale to match the pace of new threat reports, motivating the need for automated extraction methods.

Existing approaches to automated attack technique extraction fall into two broad categories.
Classification-based methods train multi-class or multi-label classifiers to predict technique identifiers directly. They struggle with severe class imbalance and data scarcity given the large label space~\cite{you2022tim,li2022attackg,nguyen2024noise,crossman2026constructing}.
Retrieval-based methods rank candidate techniques by similarity to the input text~\cite{alam2023looking,kumarasinghe2024text2ttp}, and recent hybrid pipelines combine retrieval with Large Language Model (LLM)-based generation to achieve strong results on single-label benchmarks~\cite{xu2024intelex,lekssays2025techniquerag}.
Despite this progress, a critical gap remains in the multi-label setting, where real-world CTI sentences often describe multiple techniques simultaneously.
Prompting frontier LLMs, which have broad knowledge of attack patterns, leads to over-prediction: the model infers techniques that commonly co-occur with those described in the text, even when they are not explicitly supported by the input.
Fine-tuning LLMs with token-level cross-entropy improves precision but produces conservative, single-label outputs that miss co-occurring techniques---a consequence of treating technique extraction as sequence generation rather than set prediction.
In neither case is there direct supervision on whether the predicted technique set is correct and complete.

We propose \our, a two-stage framework that addresses this gap by combining retrieval-augmented supervised fine-tuning (SFT) with reinforcement learning using verifiable rewards (RLVR).
In the first stage, a hybrid retriever narrows the large label space to a manageable candidate set, and an LLM is fine-tuned to select techniques from this set.
In the second stage, we apply Group Relative Policy Optimization (GRPO)~\cite{shao2024deepseekmath} that provides separate signals for recall, precision, and output format of the predicted technique set, giving the optimizer direct supervision that token-level objectives lack.
To handle the challenges of multi-reward optimization, we decouple the normalization of each reward channel~\cite{liu2026gdpo}, ensuring that each reward component contributes proportionally to the policy gradient.
A selective training strategy focuses RL updates on examples that remain difficult after SFT, preventing easy instances from contributing uninformative gradients.

Our contributions are as follows:
\begin{itemize}[leftmargin=*,itemsep=2pt]
  \item We formulate attack technique extraction as a \emph{retrieval-then-select} task and show that augmenting SFT with reinforcement learning can directly supervise set-level metrics, addressing the multi-label reasoning gap left by token-level training objectives.
  \item We design a decomposed verifiable reward with separate precision, recall, and format channels, combined via decoupled normalization, enabling fine-grained control over multi-label prediction quality.
  \item Across four CTI benchmarks, \our achieves the best average F1. At the sub-technique level, it outperforms Claude Sonnet 4.5 with the same RAG prompt by 7.4 percentage points, while reducing latency by 28$\times$ using an 8B model on a single GPU.
  Ablation studies validate the contribution of each component and demonstrate generalization across two base model families.
\end{itemize}
\section{Related Work}
\label{sec:related}

\begin{figure*}[t]
  \centering
  \includegraphics[width=\linewidth]{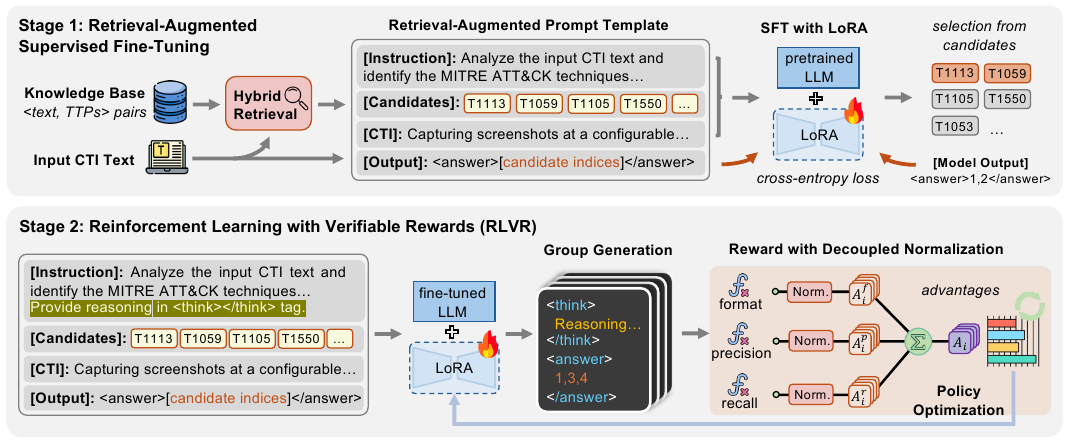}
\caption{Overview of the \our framework. Stage 1 uses a hybrid BM25 + embedding retriever to narrow the 500+ ATT\&CK taxonomy to a candidate set, then fine-tunes an LLM to select correct techniques from the candidates. Stage 2 applies GRPO with decomposed verifiable rewards that separately supervise precision, recall, and output format. Applied selectively on examples that remain difficult after SFT.
}
  \label{fig:method_overview}
\end{figure*}

\subsection{TTP Extraction from Cyber Threat Intelligence}
Mapping CTI text to MITRE ATT\&CK tactics,
techniques, and procedures (TTPs) has been approached from several angles~\cite{buchel2025sok}.
Early rule-based systems such as TTPDrill~\cite{husari2017ttpdrill} relied on hand-crafted patterns and NLP pipelines, which were precise but brittle and difficult to maintain as the ATT\&CK taxonomy evolved.
Subsequent work framed the problem as multi-class classification.
\citet{you2022tim} trained classifiers to predict technique IDs directly, but these methods struggled with severe class imbalance given the hundreds of possible labels.
Knowledge-enhanced approaches sought to mitigate this: AttacKG~\cite{li2022attackg} incorporated knowledge graphs to capture inter-technique relationships, while \citet{nguyen2024noise} applied noise-contrastive estimation to learn more discriminative technique representations.
Retrieval-based methods offer an alternative by ranking candidate techniques against the input text.
LADDER~\cite{alam2023looking} used semantic similarity for candidate ranking, and Text2TTP~\cite{kumarasinghe2024text2ttp} introduced hierarchical re-ranking with dual-encoder learning.
More recently, hybrid retrieve-then-generate pipelines have emerged.
IntelEX~\cite{xu2024intelex} combined retrieval with GPT-style zero-shot reasoning, while TechniqueRAG~\cite{lekssays2025techniquerag} anchored a fine-tuned LLM with retrieved evidence and achieved strong results on single-label benchmarks.
However, these methods optimize token-level objectives and do not explicitly address the multi-label setting where multiple techniques co-occur in a single text---a scenario that remains challenging in practice.
This gap is visible in prior results, where TechniqueRAG~\cite{lekssays2025techniquerag} performs strongly on single-label datasets but degrades substantially on the multi-label scenario.
Our work builds on the retrieve-then-select paradigm of TechniqueRAG but introduces an RL stage, specifically targeting the multi-label reasoning gap left by prior approaches.

\subsection{LLMs for Information Extraction in Security}
LLMs have shown promise across a range of security-related NLP tasks, including threat entity extraction, vulnerability analysis, and malware classification~\cite{ferrag2024generative}.
LLMCloudHunter~\cite{shabtai2024llmcloudhunter} demonstrated that LLMs can automatically extract API calls and indicators of compromise from CTI reports to generate detection rules for cloud environments.
For technique identification specifically, zero-shot prompting of frontier models such as GPT-4 has been explored~\cite{xu2024intelex}, but performance remains limited due to the domain-specific nature of the ATT\&CK taxonomy and the tendency of general-purpose LLMs to hallucinate plausible but incorrect technique IDs.
Fine-tuning on domain data substantially improves accuracy: TechniqueRAG~\cite{lekssays2025techniquerag} demonstrated that even a modest number of labelled examples suffices to adapt an instruction-tuned LLM for this task.
A common limitation of these supervised approaches is that the cross-entropy training objective treats each token independently and does not directly optimize the set-level metrics (precision, recall) that practitioners care about.
Our method addresses this by adding an RL fine-tuning stage on top of supervised fine-tuning.

\subsection{Reinforcement Learning for LLM Reasoning}
Reinforcement learning has become a key technique for aligning LLMs beyond what supervised fine-tuning alone can achieve.
RLHF~\cite{ouyang2022training} uses human preference judgements to train a reward model, but collecting such annotations is expensive and difficult to scale.
Reinforcement learning with verifiable rewards (RLVR) replaces the learned reward model with a programmatic verifier that scores outputs against ground truth.
DeepSeek-R1~\cite{guo2025deepseek} and DeepSeek-Math~\cite{shao2024deepseekmath} demonstrated that RLVR with Group Relative Policy Optimization (GRPO) can substantially improve LLM reasoning on mathematical and code tasks.
GRPO eliminates the need for a separate value network by computing advantages from multiple sampled completions per prompt, making it both simpler and more stable than standard PPO for language generation.
Rank-R1~\cite{zhuang2025rank} extended this paradigm to document re-ranking, showing that binary relevance rewards suffice to match fully supervised rerankers.
GDPO~\cite{liu2026gdpo} further improved multi-reward GRPO training by normalizing each reward channel independently before aggregation, preventing high-variance components from dominating the policy gradient.
In cybersecurity, a recent work Minerva~\cite{alam2026minerva} applies RLVR to a broad suite of CTI subtasks, including technique extraction, vulnerability mapping, and threat actor attribution. Its technique extraction predicts a single technique ID per input, rewarded by match of the technique ID, with partial credit at the base-technique level.
Our work applies RLVR with GRPO to multi-label technique extraction, where multiple techniques co-occur in a single text, and designs a decomposed reward with separate precision and recall channels combined via decoupled normalization, providing the optimizer with independent signals for correctness and completeness.

\section{Methodology}
\label{sec:method}

We present \our in two stages. Figure~\ref{fig:method_overview} gives an overview.
In Stage~1, we formulate attack technique extraction as a \emph{retrieval-then-select} task: a hybrid retriever narrows the large label space to a manageable candidate set, and a supervised fine-tuned (SFT) LLM learns to select the correct techniques from that set.
In Stage~2, we apply reinforcement learning with decomposed, verifiable rewards to directly supervise set-level correctness and completeness on examples that remain difficult after SFT.

\subsection{Stage 1: Retrieval-Augmented Supervised Fine-Tuning}
\label{sec:sft}

\noindent\textbf{Problem formulation.}
Given a CTI text $x$, the goal is to predict the set of MITRE ATT\&CK technique identifiers $Y \subseteq \mathcal{Y}$ described in $x$, where $\mathcal{Y}$ is the full taxonomy of techniques and sub-techniques.
Because $|\mathcal{Y}|$ is large (several hundred entries), directly generating technique IDs in an open-ended fashion is prone to hallucination.
We therefore decompose the problem into two steps: (i)~retrieve a small candidate set $\mathcal{T}_x \subset \mathcal{Y}$ likely to contain the correct labels, and (ii)~select the correct subset $\hat{Y} \subseteq \mathcal{T}_x$ with an LLM.

\noindent\textbf{Hybrid candidate retrieval.}
We maintain a corpus $\mathcal{C}$ of labelled CTI sentences from the training data, each associated with one or more technique identifiers.
Given input $x$, we retrieve relevant corpus entries using a hybrid of BM25~\cite{robertson2009probabilistic} and ATT\&CK-BERT~\cite{abdeen2023smet}, a Sentence-BERT~\cite{reimers2019sentence} encoder pretrained with a Siamese objective on attack-action text so that sentences describing semantically similar adversary behaviors map to nearby embeddings.
Both retrievers return ranked lists over $\mathcal{C}$ and we merge them by taking the minimum:
\begin{equation}
\mathrm{rank}(c) = \min\!\big(\mathrm{rank}_{\text{BM25}}(c),\; \mathrm{rank}_{\text{BERT}}(c)\big),
\end{equation}
and sorting accordingly.
After removing entries whose label sets are identical (deduplication), we retain the top-$k$ unique examples.
The candidate set $\mathcal{T}_x = \{t_1, \dots, t_m\}$ is formed by collecting all technique labels from these $k$ examples in retrieval-rank order, preserving first-seen ordering and capping at $m$ labels.
The hybrid design combines the lexical precision of BM25 with the semantic generalisation of dense retrieval, improving candidate recall over either method alone.

\noindent\textbf{Prompt construction and supervised fine-tuning.}
Each training example $(x, Y)$ is converted into an instruction-following sample.
The prompt comprises three parts: (i)~a system-level role description instructing the model to act as a MITRE ATT\&CK analyst, (ii)~the candidate list $\mathcal{T}_x$ with technique IDs, names, and brief descriptions, and (iii)~the CTI text $x$.
The target output enumerates the indices of the correct candidates $Y \cap \mathcal{T}_x$ within structured \texttt{<answer>} tags.
Training examples for which the retriever fails to surface any ground-truth label ($Y \cap \mathcal{T}_x = \emptyset$) are discarded, as they provide no useful supervision signal.
We fine-tune a pre-trained instruction-tuned LLM on these samples with Low-Rank Adaptation (LoRA)~\cite{hu2022lora} by minimizing the standard causal language modelling loss:
\begin{equation}
  \mathcal{L}_{\text{SFT}}(\theta) = -\mathbb{E}_{(x,\,Y)}\!\left[\sum_{t=1}^{|y^*|} \log \pi_\theta\!\big(y^*_t \mid y^*_{<t},\, x,\, \mathcal{T}_x\big)\right],
  \label{eq:sft}
\end{equation}
where $y^* = \texttt{<answer>}\, \{i : t_i \in Y \cap \mathcal{T}_x\} \,\texttt{</answer>}$ is the target sequence.
This produces the SFT policy $\pi_{\text{SFT}}$.
Figure~\ref{fig:sft_prompt} shows the complete prompt template.

\begin{figure}[t]
\centering
\small
\begin{tcolorbox}[colback=blue!3, colframe=blue!40, title={\textbf{SFT Prompt Template}}, fonttitle=\small]
\texttt{\textbf{[User]}} \\
You are a cybersecurity expert specializing in the MITRE ATT\&CK framework. Your task is to analyze cyber threat descriptions and identify the associated MITRE ATT\&CK techniques. The MITRE ATT\&CK framework is a globally-accessible knowledge base of adversary tactics and techniques based on real-world observations. This framework is used for developing specific threat models and methodologies in the private sector, government, and cybersecurity product and service community. Analyze the input cyber threat description and list the relevant MITRE ATT\&CK techniques. Select the MITRE ATT\&CK techniques from the following list and output their number in \texttt{<answer> </answer>} tags. \\[4pt]
A list of techniques that may be relevant: \\
\texttt{[1] T1547.001 Boot or Logon Autostart Execution: {\color{gray}\textit{[description]}}} \\
\texttt{[2] T1059.001 Command and Scripting Interpreter: {\color{gray}\textit{[description]}}} \\
\texttt{{\color{gray}\textit{[...$m$ candidates total...]}}} \\[4pt]
Input cyber threat description: \\
{\color{gray}\textit{[CTI text $x$]}} \\[6pt]
\texttt{\textbf{[Assistant]}} \\
\texttt{<answer>} \\
\texttt{1} \\
\texttt{</answer>}
\end{tcolorbox}
\caption{SFT prompt template used in Stage 1 training. The model receives a cybersecurity expert role description, a CTI text passage, and a numbered candidate list populated by the hybrid retriever. The target output selects the correct candidates by their indices within <answer> tags.}
\label{fig:sft_prompt}
\end{figure}

\subsection{Stage 2: Reinforcement Learning with Verifiable Rewards}
\label{sec:rlvr}

The SFT objective optimizes token-level cross-entropy, which does not directly correspond to set-level evaluation metrics such as precision and recall.
It penalises the model for generating correct techniques in the wrong order and offers no direct mechanism to supervise whether the predicted set is correct and complete; in practice, this causes the SFT model to favour conservative, single-label predictions that achieve high precision but miss ground-truth techniques.
To bridge this gap, we further train $\pi_{\text{SFT}}$ with reinforcement learning using rewards that are automatically computed from the ground-truth labels.

\noindent\textbf{Selective training.}
We first run inference with $\pi_{\text{SFT}}$ on the training set multiple times and measure per-example F1.
Only examples whose average F1 falls below a threshold $\tau$ are retained for RL training.
This focuses optimization on the cases where the SFT model still produces errors, avoiding redundant gradient updates on already-solved instances and improving sample efficiency.

\noindent\textbf{Structured generation.}
During RL, the model is prompted to first produce a free-form reasoning trace enclosed in \texttt{<think>}\,\ldots\,\texttt{</think>} tags, followed by its final predictions in an \texttt{<answer>}\,\ldots\,\texttt{</answer>} block.
Only the content of the \texttt{<answer>} block is used for reward computation; the reasoning trace serves as a chain-of-thought scaffold~\cite{wei2022chain} that empirically improves answer quality without being directly supervised.
RL prompt template is shown in Figure~\ref{fig:rl_prompt}.

\begin{figure}[t]
\centering
\small
\begin{tcolorbox}[colback=blue!3, colframe=blue!40, title={\textbf{RL Prompt Template} (additions over SFT prompt highlighted in \textcolor{blue}{blue})}, fonttitle=\small]
\texttt{\textbf{[User]}} \\
You are a cybersecurity expert specializing in the MITRE ATT\&CK framework. Your task is to analyze cyber threat descriptions and identify the associated MITRE ATT\&CK techniques. {\color{gray}\textit{[...framework description...]}} Analyze the input cyber threat description and list the relevant MITRE ATT\&CK techniques. \textcolor{blue}{Provide your reasoning process within \texttt{<think> </think>} tags.} Select the MITRE ATT\&CK techniques from the following list and output their number in \texttt{<answer> </answer>} tags. \\[4pt]
A list of techniques that may be relevant: \\
\texttt{[1] T1547.001 Boot or Logon Autostart Execution: {\color{gray}\textit{[description]}}} \\
\texttt{[2] T1059.001 Command and Scripting Interpreter: {\color{gray}\textit{[description]}}} \\
\texttt{{\color{gray}\textit{[...$m$ candidates total...]}}} \\[4pt]
Input cyber threat description: \\
{\color{gray}\textit{[CTI text $x$]}}
\end{tcolorbox}
\caption{RL prompt template used in Stage 2 training. Compared to the SFT prompt (Figure~\ref{fig:sft_prompt}), the model is additionally instructed to produce a reasoning trace within <think> tags before committing to its answer.}
\label{fig:rl_prompt}
\end{figure}

\noindent\textbf{Reward design.}
Given a completion, we extract the predicted set $\hat{Y}$ from the answer block and compute three reward signals against the ground truth $Y$:

\begin{itemize}[leftmargin=*,itemsep=2pt]
  \item \textbf{Precision reward} $r_P$\textbf{.}\; We evaluate precision at both the technique level and the sub-technique level, and average the two.
  \begin{equation}
    r_P = \tfrac{1}{2}\!\left(\frac{|\hat{Y}_T \cap Y_T|}{|\hat{Y}_T|} + \frac{|\hat{Y}_S \cap Y_S|}{|\hat{Y}_S|}\right),
  \end{equation}
  where subscripts $T$ and $S$ denote parent-technique and sub-technique granularities, respectively.
  \item \textbf{Recall reward} $r_R$\textbf{.}\; Similarly, we evaluate recall at both technique and the sub-technique level, and take the average. Defined analogously over ground-truth set sizes:
  \begin{equation}
    r_R = \tfrac{1}{2}\!\left(\frac{|\hat{Y}_T \cap Y_T|}{|Y_T|} + \frac{|\hat{Y}_S \cap Y_S|}{|Y_S|}\right).
  \end{equation}
  \item \textbf{Format reward} $r_F$\textbf{.}\; A discrete score that verifies whether the output contains well-formed \texttt{<think>} and \texttt{<answer>} blocks with non-empty content, encouraging adherence to the structured output format.
\end{itemize}

The final reward is a weighted sum: 
\begin{equation}
r = w_P\, r_P + w_R\, r_R + w_F\, r_F.
\end{equation}
A key design choice is to \emph{decompose} precision and recall into separate reward channels rather than collapsing them into a single F1 score.
A single F1 reward provides only one scalar signal per completion, making it difficult for the optimizer to distinguish whether a low score stems from over-prediction (low precision) or under-prediction (low recall).
By decomposing, the optimizer receives independent gradient signals for each axis, enabling finer control over set-level metrics through the weights $w_P$ and $w_R$.
Moreover, evaluating at both granularities simultaneously encourages the model to predict the correct parent technique even when it is uncertain about the specific sub-technique.

\noindent\textbf{Policy optimization with reward-decoupled normalization.}
We optimize the policy $\pi_\theta$ (initialized from $\pi_{\text{SFT}}$) using Group Relative Policy Optimization (GRPO)~\cite{shao2024deepseekmath}.
For each prompt $x$, it samples a group of $G$ completions from $\pi_\theta$.
In standard GRPO, the advantage for the $i$-th completion is $A_i = (r_i - \mu) / \sigma$, where $\mu$ and $\sigma$ are the mean and standard deviation of the scalar rewards within the group.
When multiple reward channels are combined into a single scalar before normalization, distinct reward combinations can collapse into identical advantage values, losing the training signal that distinguishes them.
Inspired by~\citet{liu2026gdpo}, we address this by decoupling the normalization across reward channels: each channel is normalized independently within the group before weighted aggregation.
Let $r_i^{(j)}$ denote the $j$-th reward channel for completion $i$, and let $\mu^{(j)}, \sigma^{(j)}$ be the within-group mean and standard deviation of that channel.
The reward-decoupled advantage is:
\begin{equation}
  \hat{A}_i = \mathrm{BN}\!\left(\sum_{j} w_j \cdot \frac{r_i^{(j)} - \mu^{(j)}}{\sigma^{(j)} + \delta}\right),
  \label{eq:advantage}
\end{equation}
where $\text{BN}(\cdot)$ denotes batch-level mean-variance normalization and $\delta$ is a small constant for numerical stability.
This preserves the relative differences across reward channels regardless of their raw scale or variance.

The full training objective is:
\begin{equation}
\begin{split}
\mathcal{L}(\theta) = -\mathbb{E}_{x}\Bigg[&\frac{1}{G}\sum_{i=1}^{G} \min\!\Big(\rho_i\,\hat{A}_i,\;\mathrm{clip}(\rho_i, 1{-}\epsilon, 1{+}\epsilon)\,\hat{A}_i\Big) \\
&- \beta\, D_{\mathrm{KL}}\!\big(\pi_\theta \| \pi_{\text{SFT}}\big)\Bigg],
\end{split}
\label{eq:grpo}
\end{equation}
where $\rho_i = \pi_\theta(y_i \mid x) / \pi_{\mathrm{old}}(y_i \mid x)$ is the importance-sampling ratio, $\epsilon$ controls ratio clipping, and $\beta$ weights a KL penalty that prevents the policy from drifting too far from the SFT initialization.
After RL training, the LoRA adapter is merged into the base model to produce the final policy $\pi_{\text{\our}}$.

\section{Experiments}
\label{sec:experiments}

\begin{table}[t]
\centering
\caption{Statistics for the four CTI benchmarks.}
\label{tab:dataset-stats}
\scalebox{0.77}{
\begin{tabular}{lcc cc cc cc}
\toprule
\multirow{2}{*}{Dataset} & \multicolumn{2}{c}{\# Samples} & \multicolumn{2}{c}{Avg.\ Labels} & \multicolumn{2}{c}{Avg.\ Words} & \multicolumn{2}{c}{Unique TTPs} \\
\cmidrule(lr){2-3} \cmidrule(lr){4-5} \cmidrule(lr){6-7} \cmidrule(lr){8-9}
& Train & Test & Train & Test & Train & Test & Train & Test \\
\midrule
TRAM & 4{,}072 & 725 & 1.16 & 1.17 & 21.0 & 21.3 & 184 & 127 \\
Procedures & 9{,}956 & 1{,}767 & 1.01 & 1.00 & 13.4 & 13.4 & 474 & 300 \\
Derived Procedures & 2{,}998 & 521 & 1.23 & 1.20 & 67.3 & 64.9 & 359 & 194 \\
Expert & 540 & 157 & 1.42 & 3.32 & 38.2 & 71.9 & 251 & 152 \\
\bottomrule
\end{tabular}
}
\end{table}

\subsection{Datasets}
We evaluate on four benchmarks\footnote{Datasets are available at: \url{https://github.com/tumeteor/mitre-ttp-mapping}} for MITRE ATT\&CK technique annotation from CTI text, adopting the same splits as~\citet{lekssays2025techniquerag}. We use the combined train and dev splits for fine-tuning. Table~\ref{tab:dataset-stats} summarizes the statistics. The descriptions are as follows:
\begin{itemize}[leftmargin=*,itemsep=4pt]
\item \textbf{TRAM}~\cite{tram2023} is a large publicly available manually curated dataset, released by the Center for Threat-Informed Defense. It comprises short text fragments extracted from threat reports, each labeled with ATT\&CK (sub-)technique IDs. As noted by~\citet{nguyen2024noise}, TRAM covers only about one-third of the ATT\&CK taxonomy and contains relatively noisy labels. The fragments often lack surrounding context (e.g., ``can download and execute a file from given URL'' with no subject indicating what tool or malware performs the action), making it difficult for technique assignment.

\item \textbf{Procedures}~\cite{nguyen2024noise} consists of one-sentence expert-written summaries of how a technique is implemented in real-world attacks, collected from the ATT\&CK knowledge base (v12.0) procedure examples. This is effectively a single-label dataset. With 300 unique sub-techniques in the test split, it tests breadth of coverage but is relatively straightforward since sentences closely mirror technique definitions.

\item \textbf{Derived Procedures}~\cite{nguyen2024noise} complements the Procedures dataset with longer passages that align to threat report writing style. These are evidential paragraphs sourced from the URL references cited in ATT\&CK procedure examples, identified via per-document search. Unlike Procedures, these passages require reasoning over implicit adversary behaviors rather than matching explicit technique descriptions.

\item \textbf{Expert}~\cite{nguyen2024noise} is a paragraph-level dataset annotated by five CTI experts, designed to closely emulate real-world technique extraction scenarios. Unlike the sentence-focused datasets above, Expert covers entire paragraphs, and annotations are inherently multi-label (avg.\ 3.32 labels, up to 18 in the test split). 
This is the most challenging benchmark, testing the model's ability to identify multiple co-occurring techniques.
\end{itemize}

\begin{table*}[t]
\centering
\caption{Technique-level evaluation (\%). Sub-technique IDs are truncated to their parent before computing metrics. Best results in \textbf{bold}, second-best underlined. Avg.\ macro-F1 and Rank are computed across all four datasets. \our achieves the highest average F1 and the best average rank, outperforming frontier LLMs with RAG and domain-specific baselines.}
\label{tab:main_technique}
\begin{tabular}{@{}l ccc ccc ccc ccc | cc@{}}
\toprule
\multirow{2}{*}[2pt]{\textbf{Model}} & \multicolumn{3}{c}{\textbf{TRAM}} & \multicolumn{3}{c}{\textbf{Procedures}} & \multicolumn{3}{c}{\textbf{Derived Procedures}} & \multicolumn{3}{c}{\textbf{Expert}} & \multicolumn{2}{|c}{\textbf{Avg.}} \\
\cmidrule(lr){2-4} \cmidrule(lr){5-7} \cmidrule(lr){8-10} \cmidrule(lr){11-13} \cmidrule(lr){14-15}
& Prec. & Rec. & F1 & Prec. & Rec. & F1 & Prec. & Rec. & F1 & Prec. & Rec. & F1 & F1 & Rank \\
\midrule
DeepSeek-R1          & 49.2 & 72.8 & 58.7 & 64.4 & 85.9 & 73.6 & 28.0 & 66.8 & 39.4 & 43.6 & 59.1 & 50.2 & 55.5 & 7.5 \\
\quad w/ RAG         & 54.7 & 80.0 & 65.0 & 68.5 & \underline{91.9} & 78.5 & 31.1 & 71.6 & 43.3 & 45.9 & 59.3 & 51.8 & 59.7 & 4.5 \\
Claude Sonnet 4.5    & 52.9 & 76.3 & 62.5 & 67.6 & 84.8 & 75.2 & 28.2 & 72.6 & 40.6 & 46.5 & 67.5 & \underline{55.1} & 58.3 & 5.5 \\
\quad w/ RAG         & 59.4 & 78.8 & 67.8 & 73.7 & \textbf{92.9} & 82.2 & 31.8 & \underline{73.6} & 44.4 & 59.0 & \underline{70.3} & \textbf{64.2} & 64.7 & \underline{3.2} \\
Qwen3-8B             & 20.4 & 28.6 & 23.9 & 25.0 & 32.2 & 28.1 & 12.6 & 29.0 & 17.5 & 20.2 & 25.6 & 22.6 & 23.0 & 12.5 \\
\quad w/ RAG         & 52.3 & 73.3 & 61.0 & 67.0 & 85.9 & 75.3 & 32.4 & 58.5 & 41.7 & 47.1 & 50.1 & 48.6 & 56.6 & 6.5 \\
Ministral-8B         & 13.5 & 26.1 & 17.8 & 16.0 & 28.6 & 20.5 & 8.5  & 21.9 & 12.2 & 11.0 & 15.5 & 12.9 & 15.8 & 14.0 \\
\quad w/ RAG         & 46.8 & 78.2 & 58.6 & 53.3 & 79.7 & 63.9 & 26.3 & 71.0 & 38.4 & 31.0 & 50.5 & 38.4 & 49.8 & 10.2 \\
Nova 2 Lite          & 20.6 & 48.2 & 28.9 & 26.7 & 54.9 & 35.9 & 8.2  & 34.8 & 13.3 & 13.8 & 33.4 & 19.6 & 24.4 & 12.5 \\
\quad w/ RAG         & 51.6 & 71.5 & 59.9 & 62.0 & 79.8 & 69.8 & 28.9 & 61.0 & 39.3 & 45.6 & 48.3 & 46.9 & 54.0 & 8.5 \\
IntelEX              & 35.9 & \textbf{85.4} & 50.6 & 47.0 & 91.4 & 62.1 & 16.0 & \textbf{82.4} & 26.7 & 32.5 & \textbf{82.9} & 46.7 & 46.5 & 10.5 \\
TechniqueRAG         & 76.0 & 72.1 & 74.0 & \textbf{91.1} & 91.1 & \textbf{91.1} & 46.8 & 42.9 & 44.8 & \textbf{75.2} & 37.7 & 50.2 & 65.0 & \underline{3.2} \\
\midrule
\our w/o RL          & \textbf{82.2} & 79.2 & \underline{80.6} & \underline{89.9} & 89.8 & 89.9 & \textbf{52.7} & 49.0 & \underline{50.8} & \underline{67.8} & 33.6 & 45.0 & \underline{66.6} & 4.2 \\
\our                 & \underline{80.2} & \underline{84.5} & \textbf{82.3} & 89.5 & 90.5 & \underline{90.0} & \underline{48.1} & 57.8 & \textbf{52.5} & 64.8 & 41.5 & 50.6 & \textbf{68.9} & \textbf{2.0} \\
\bottomrule
\end{tabular}
\end{table*}

\begin{table*}[t]
\centering
\caption{Sub-technique-level evaluation (\%). Predictions must match the full sub-technique ID. \our's advantage grows at this finer granularity (+7.4 F1 over Claude Sonnet 4.5 w/ RAG, vs.\ +4.2 at technique-level).}
\label{tab:main_subtechnique}
\begin{tabular}{@{}l ccc ccc ccc ccc | cc@{}}
\toprule
\multirow{2}{*}[2pt]{\textbf{Model}} & \multicolumn{3}{c}{\textbf{TRAM}} & \multicolumn{3}{c}{\textbf{Procedures}} & \multicolumn{3}{c}{\textbf{Derived Procedures}} & \multicolumn{3}{c}{\textbf{Expert}} & \multicolumn{2}{|c}{\textbf{Avg.}} \\
\cmidrule(lr){2-4} \cmidrule(lr){5-7} \cmidrule(lr){8-10} \cmidrule(lr){11-13} \cmidrule(lr){14-15}
& Prec. & Rec. & F1 & Prec. & Rec. & F1 & Prec. & Rec. & F1 & Prec. & Rec. & F1 & F1 & Rank \\
\midrule
DeepSeek-R1          & 36.2 & 57.8 & 44.6 & 55.7 & 76.3 & 64.4 & 21.6 & 54.9 & 31.0 & 34.4 & 44.6 & 38.8 & 44.7 & 9.2 \\
\quad w/ RAG         & 41.2 & 68.7 & 51.5 & 61.1 & 87.7 & 72.0 & 25.7 & 68.1 & 37.3 & 41.2 & 50.3 & 45.3 & 51.5 & 4.5 \\
Claude Sonnet 4.5    & 43.2 & 67.5 & 52.7 & 59.8 & 79.9 & 68.4 & 22.2 & 64.4 & 33.0 & 43.0 & 57.7 & \underline{49.3} & 50.9 & 5.5 \\
\quad w/ RAG         & 48.8 & 76.0 & 59.4 & 61.4 & \textbf{88.9} & 72.6 & 25.1 & 69.2 & 36.8 & 53.8 & 61.0 & \textbf{57.2} & 56.5 & 3.5 \\
Qwen3-8B             & 14.7 & 23.4 & 18.1 & 12.3 & 16.9 & 14.2 & 5.6  & 14.9 & 8.2  & 15.6 & 18.4 & 16.9 & 14.3 & 12.5 \\
\quad w/ RAG         & 38.9 & 61.8 & 47.7 & 60.2 & 81.2 & 69.1 & 27.0 & 53.2 & 35.8 & 41.6 & 42.3 & 42.0 & 48.6 & 6.8 \\
Ministral-8B         & 8.6  & 20.8 & 12.2 & 6.4  & 14.7 & 8.9  & 3.1  & 11.3 & 4.8  & 8.3  & 9.8  & 9.0  & 8.7  & 14.0 \\
\quad w/ RAG         & 36.4 & 69.0 & 47.7 & 44.0 & 71.7 & 54.5 & 21.2 & 64.5 & 31.9 & 26.8 & 41.6 & 32.6 & 41.7 & 9.8 \\
Nova 2 Lite          & 15.9 & 40.8 & 22.9 & 16.1 & 38.0 & 22.6 & 3.8  & 20.5 & 6.4  & 10.5 & 24.4 & 14.7 & 16.6 & 12.5 \\
\quad w/ RAG         & 41.6 & 65.0 & 50.7 & 53.3 & 76.0 & 62.7 & 24.0 & 56.3 & 33.6 & 40.6 & 41.4 & 41.0 & 47.0 & 7.8 \\
IntelEX              & 28.3 & \textbf{83.3} & 42.2 & 35.5 & 87.0 & 50.4 & 12.2 & \textbf{79.8} & 21.2 & 29.8 & \textbf{79.7} & 43.4 & 39.3 & 9.5 \\
TechniqueRAG         & 72.7 & 68.7 & 70.7 & \textbf{91.1} & \underline{88.1} & \textbf{88.1} & 40.1 & 35.9 & 37.9 & \textbf{70.1} & 30.2 & 42.2 & 59.7 & \underline{3.2} \\
\midrule
\our w/o RL          & \textbf{78.8} & 75.9 & \underline{77.3} & \underline{85.7} & 85.7 & 85.7 & \textbf{46.9} & 44.0 & \underline{45.4} & \underline{65.0} & 27.3 & 38.5 & \underline{61.7} & 4.2 \\
\our                 & \underline{76.4} & \underline{81.3} & \textbf{78.7} & 85.1 & 86.5 & \underline{85.8} & \underline{42.0} & 52.6 & \textbf{46.7} & 60.8 & 34.7 & 44.2 & \textbf{63.9} & \textbf{2.0} \\
\bottomrule
\end{tabular}
\end{table*}

\subsection{Compared Methods}
We compare \our against three categories of baselines: 
\begin{enumerate}[leftmargin=*,itemsep=4pt]
\item{\textbf{Frontier LLMs.}}
We evaluate DeepSeek-R1~\cite{guo2025deepseek} and Claude Sonnet 4.5~\cite{anthropic2025sonnet45}, both in zero-shot and RAG-augmented settings. These represent the strong general-purpose reasoning models available at the time of evaluation. In the zero-shot setting, models receive only the task instruction and input text. In the RAG setting, we provide the same retrieved candidate list used by \our---up to 25 techniques with their IDs and short descriptions---in the prompt, asking the model to select from among them.

\item{\textbf{Open-weight LLMs.}}
We evaluate Qwen3-8B~\cite{yang2025qwen3technicalreport}, Ministral-8B-Instruct~\cite{mistral2024ministral}, and Amazon Nova 2 Lite~\cite{Intelligence2025}, each in zero-shot and RAG-augmented settings. These models match \our in parameter count (8B) and serve as direct comparisons to isolate the effect of domain-specific fine-tuning from model scale. Ministral-8B is the same base model used by \our and TechniqueRAG, enabling controlled comparison.

\item{\textbf{Domain-Specific methods.}}
IntelEX~\cite{xu2024intelex} is a training-free pipeline that combines retrieval-augmented candidate generation with LLM-as-a-judge validation. For each candidate technique retrieved from the MITRE knowledge base, an LLM independently determines whether it is present in the input text by generating a YES/NO judgment with justification. We re-implement IntelEX using Claude Sonnet 4.5 as the backbone LLM (the original uses GPT-4o-mini). The pipeline validates all RAG candidates individually, which yields high recall but low precision due to the inability to jointly reason about which subset of candidates is most relevant.
TechniqueRAG~\cite{lekssays2025techniquerag} is the prior state-of-the-art framework for this task. It retrieves candidate text--technique pairs via BM25, re-ranks them using DeepSeek V3 with a domain-specific prompting framework that decomposes queries into attack steps and evaluates sub-technique relevance, then fine-tunes Ministral-8B as the generator on the top re-ranked few-shot examples. We report their published results on the same test splits. Since TechniqueRAG does not have results on Derived Procedures, we run their released model on this dataset using the same evaluation protocol for a complete comparison.
\end{enumerate}
For all models in (1) and (2), we show the performance both without and with RAG. In the RAG setting, we provide the same retrieved candidate list (up to 25 techniques) used by \our.

\subsection{Implementation Details}
\noindent\textbf{Retrieval.}
We use BM25 (Okapi, $k_1{=}1.6$, $b{=}0.75$) combined with ATT\&CK-BERT~\cite{abdeen2023smet}, a domain-pretrained Sentence-BERT encoder, and FAISS for dense retrieval. We retrieve $k{=}25$ candidate techniques per query, presenting each as a numbered option with its technique ID and description.
The retrieval recall at the sub-technique level is 97.3\% on TRAM, 96.9\% on Procedures, 89.7\% on Derived Procedures, and 78.6\% on Expert. As the LLM selects only from the retrieved candidate list, this caps the model's attainable recall.
Training examples for which no correct technique appears in the candidate list provide no useful supervision and are discarded before SFT (786 out of 17,566), resulting in 16{,}780 training samples.

\noindent\textbf{Training.}
We fine-tune Ministral-8B-Instruct~\cite{mistral2024ministral} in two stages. 
\begin{itemize}[leftmargin=*,itemsep=4pt]
\item\textbf{SFT}: LoRA ($r{=}32$, $\alpha{=}64$) on 16{,}780 training samples for 3 epochs with learning rate $10^{-4}$ and effective batch size 64. 
\item\textbf{RL}: Initialized from the SFT checkpoint, we apply GRPO~\cite{shao2024deepseekmath} with LoRA ($r{=}8$, $\alpha{=}16$) for 1 epoch. We sample $G{=}16$ completions per prompt at temperature $1.4$ and optimize a reward combining precision and recall: $r = 0.45P + 0.45R + 0.1r_F$. A selective training threshold $\tau{=}0.99$ skips prompts where all completions already achieve near-perfect F$_1$. 
\end{itemize}
All training uses DeepSpeed ZeRO-2 on a \texttt{g6e.48xlarge} instance with $8{\times}$ NVIDIA L40S GPUs.

\noindent\textbf{Inference.}
For inference, all models decode with temperature $0.7$ and top-$p$ $0.1$, generating up to 1{,}024 tokens. DeepSeek-R1, Claude Sonnet 4.5, and Nova 2 Lite are served via Amazon Bedrock; Qwen3-8B, Ministral-8B, and our models are run locally. Latency for \our is measured on a single NVIDIA L40S GPU with unoptimized HuggingFace inference.

\subsection{Evaluation Metrics}
We report sample-level Precision, Recall, and F1 over predicted vs.\ ground-truth technique ID sets, macro-averaged across all test samples. For each test sample, let $P_i$ be the set of predicted technique IDs and $G_i$ the set of ground-truth IDs. We compute:
\begin{align*}
&\text{Precision}_i = \frac{|P_i \cap G_i|}{|P_i|}, \\ %\quad
&\text{Recall}_i = \frac{|P_i \cap G_i|}{|G_i|}, \\
&\text{F}_1{}_i = \frac{2 \cdot \text{Precision}_i \cdot \text{Recall}_i}{\text{Precision}_i + \text{Recall}_i},
\end{align*}
Final metrics are macro-averaged: $\text{F}_1 = \frac{1}{N}\sum_{i=1}^{N} \text{F}_1{}_i$. 
We evaluate at two granularities: technique-level (parent IDs, e.g., T1027) and sub-technique-level (full IDs, e.g., T1027.002). 
At the sub-technique level, a prediction is correct only if it exactly matches a ground-truth full technique ID. Predicting a parent technique when a sub-technique is the gold label, or vice versa, is not considered correct.
At the technique level, all sub-technique IDs are first truncated to their parent IDs (e.g., T1027.002 $\rightarrow$ T1027) before computing metrics.
We also report average F1 across all four datasets and average rank.

\subsection{Main Results}

Tables~\ref{tab:main_technique} and~\ref{tab:main_subtechnique} present results at the technique and sub-technique levels, respectively.
\our achieves the highest average F1 at both granularities and the best average rank across all four datasets. 
We highlight three key findings.

\noindent\textbf{Non-fine-tuned models over-predict while SFT models under-predict.}
Non-fine-tuned models tend to over-predict, achieving high recall at the expense of precision.
For example, Claude Sonnet 4.5 attains recall above 67\% on all datasets yet precision drops as low as 28\% on Derived Procedures.
It predicts on average 2 labels per sample on single-label benchmarks and 4.2 on multi-label benchmarks (ground-truth average 1.7).
Fine-tuned models exhibit the opposite pattern.
TechniqueRAG achieves the highest precision among all methods on Expert (75.2\%) but recall drops to just 37.7\%.
It predicts only 1.01 labels per sample on average---including Expert where the ground truth averages 3.32---suggesting that token-level fine-tuning drives the model toward conservative, single-label predictions that miss co-occurring techniques.
This motivates our RL stage: rather than optimizing token-level likelihood, which lacks direct control over set-level metrics, decomposed rewards allow the optimizer to recover recall from the conservative SFT policy while preserving its precision gains.

\noindent\textbf{Domain-specific fine-tuning closes the gap between small LMs and frontier models.}
Without fine-tuning, raw 8B-class models perform poorly, far below frontier LLMs.
RAG narrows this gap substantially, with disproportionate gains for smaller models: 8B-class LMs improve by 121--215\% in average F1 at the technique level, compared to 8--11\% for frontier models.
This indicates that a constrained candidate set is most beneficial when the base model lacks domain knowledge, transforming open-ended generation into a manageable selection task.
SFT pushes performance further still, surpassing Claude Sonnet 4.5 w/ RAG. This suggests retrieval with structured fine-tuning is more effective than scaling model size alone.

\noindent\textbf{RL improves F1 by selectively boosting recall.}
RL achieves relative gains of 3.4\% at both granularities over the SFT baseline (i.e., \our w/o RL), with improvements concentrated on the more challenging multi-label datasets.
Recall increases by 18\% on Derived Procedures and 24\% on Expert at technique-level, while Procedures dataset remains stably high performance.
This comes with a modest precision trade-off, but the net F1 effect is consistently positive---most notably improving by 12\% on Expert.

\subsection{Efficiency Analysis}

\begin{figure}[t]
  \centering
  \includegraphics[width=0.88\linewidth]{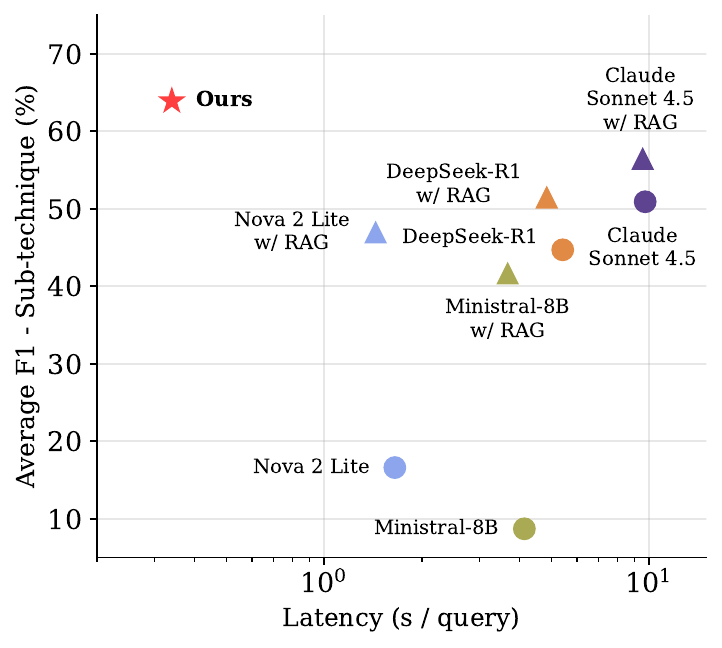}
  \caption{Per-query latency (seconds) vs.\ average sub-technique F1. \our achieves the highest F1 at 0.34s per query---28$\times$ faster than Claude Sonnet 4.5.}
  \label{fig:latency_performance}
\end{figure}

Figure~\ref{fig:latency_performance} shows per-query latency against average sub-technique F1.
Baseline latencies are measured via Amazon Bedrock API calls, which benefit from Bedrock's optimized serving infrastructure.
\our is measured locally on a single L40S GPU with unoptimized HuggingFace inference---a conservative estimate, as production serving (e.g., vLLM) would further reduce latency.

\our achieves the highest F1 at 0.34s per query. This is 4$\times$ faster than Nova 2 Lite (1.44s), 14$\times$ faster than DeepSeek-R1 (4.84s), and 28$\times$ faster than Claude Sonnet 4.5 (9.55s).
The gap is driven by output length. Reasoning models like DeepSeek-R1 produce long chain-of-thought traces before arriving at an answer, and Claude Sonnet 4.5 generates detailed per-technique justifications, both yielding hundreds of output tokens.
Fine-tuning teaches \our to directly emit technique indices in ${\sim}$9 tokens on average, and since autoregressive decoding scales linearly with sequence length, this brevity translates directly into the observed speedup.

\begin{figure}[t]
  \centering
  \includegraphics[width=1.\linewidth]{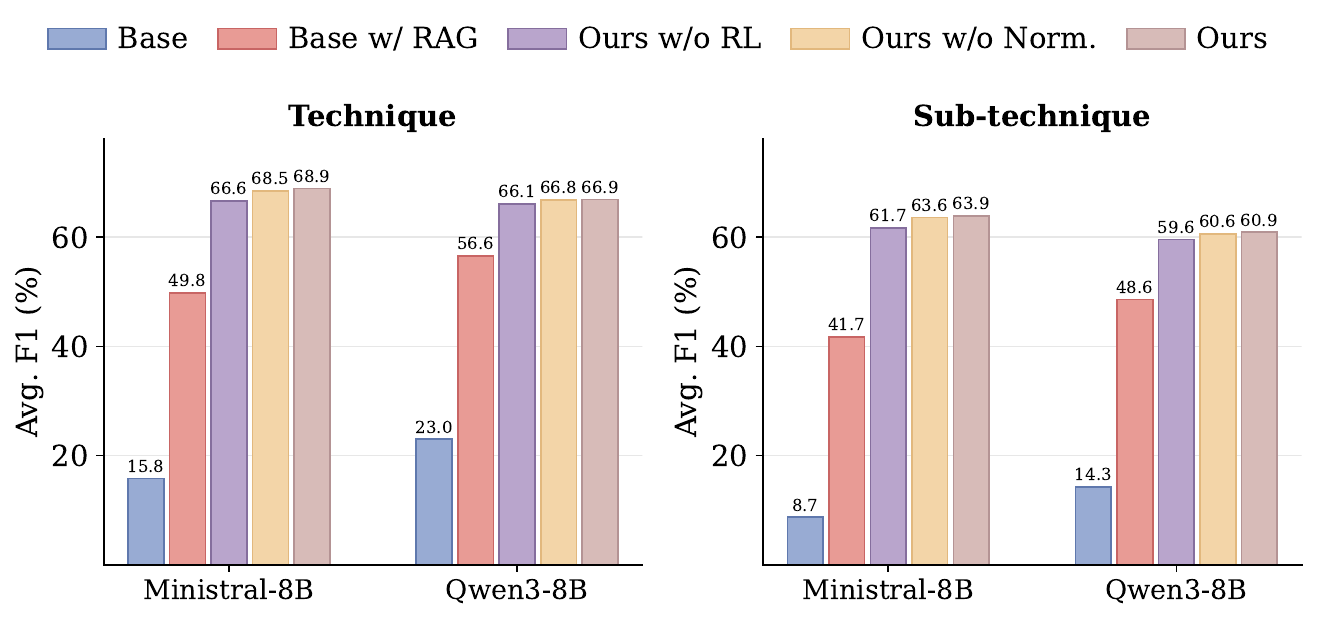}
\caption{Ablation study across two base models (Ministral-8B and Qwen3-8B). RAG yields the largest gain, SFT the second-largest, and RL adds a targeted boost concentrated on multi-label benchmarks (see Table~\ref{tab:main_technique} and~\ref{tab:main_subtechnique}).}
  \label{fig:ablation}
\end{figure}

\subsection{Ablation Study}

Figure~\ref{fig:ablation} presents the ablation study across two base models (Ministral-8B and Qwen3-8B).
The largest gain comes from using RAG, which constrains the output space from hundreds of techniques to a manageable candidate list.
SFT contributes the second-largest improvement by teaching the model to reliably parse candidates and produce structured outputs.
RL adds a smaller but targeted gain, concentrated on multi-label examples (e.g., +5.6 F1 on Expert for Ministral-8B) without degrading easier benchmarks.
Removing decoupled reward normalization reduces Ministral-8B by 0.4 F1, suggesting the benefit of it stabilizing training by preventing high-variance reward components from dominating the gradient.
Notably, both base models converge to similar final performance, despite Qwen3-8B starting substantially higher in the zero-shot setting. This suggests that the pipeline's effectiveness is driven more by the retrieval and training stages than by the base model's prior knowledge.

\section{Conclusion}
We presented \our, a two-stage framework for multi-label attack technique extraction from cyber threat intelligence text.
By combining retrieval-augmented supervised fine-tuning with reinforcement learning using decomposed verifiable rewards, \our directly optimizes set-level correctness and completeness---addressing a fundamental limitation of token-level training objectives used by prior methods.
Across four CTI benchmarks, \our achieves state-of-the-art average F1 at both technique and sub-technique granularities, with particularly large gains on recall-demanding multi-label datasets.
The approach is efficient: an 8B-parameter model running on a single GPU achieves up to 28$\times$ lower latency than frontier LLM baselines while delivering superior average performance.
Ablation studies confirm that each component---retrieval, SFT, RL, and decoupled normalization---contributes to the final result, and the pipeline generalizes across two base model families.

\noindent\textbf{Limitations and future work.}
Our approach has several limitations. 
First, the retriever imposes an upper bound on recall: if the correct technique is not in the candidate set, the model cannot predict it. Improving retrieval coverage, e.g., through query expansion or iterative retrieval, could address this. 
Second, the RL gains, while consistent, are modest compared to the SFT stage, suggesting that the current reward design may not fully exploit the potential of policy optimisation.
Third, our evaluation is limited to sentence- and paragraph-level extraction; scaling to full-document TTP extraction remains an open challenge due to lack of labeled benchmark.
Promising directions include incorporating analyst-in-the-loop feedback to refine the reward signal on ambiguous cases, extending the framework to full-document TTP extraction, and adapting the approach to related or evolving taxonomies (e.g., new ATT\&CK versions or MITRE ATLAS~\cite{mitreatlas} for AI system threats) via few-shot updates~\cite{zhang2025react}.

\bibliographystyle{ACM-Reference-Format}
\bibliography{ref}

\end{document}